\documentclass[preprint,12pt]{elsarticle}
\usepackage{caption}

\makeatletter
\def\ps@pprintTitle{%
   \let\@oddhead\@empty
   \let\@evenhead\@empty
   \let\@oddfoot\@empty
   \let\@evenfoot\@oddfoot
}
\makeatother

\usepackage{amssymb}
\usepackage{braket}
\usepackage{amsthm}
\usepackage{mathtools}
\usepackage{physics}
\usepackage{hyperref}
\usepackage[usenames, dvipsnames]{color}

\usepackage{amssymb,mathtools}
\def\eq#1{{Eq.~(\ref{#1})}}

\def\fig#1{{Fig.~\ref{#1}}}

\def\sect#1{{Section~\ref{#1}}}

\def\abs#1{\left| #1\right|}

\def\Im{\mbox{Im}\,}
\def\Re{\mbox{Re}\,}

\begin{document}

\begin{frontmatter}

\title{$B_s$-$\bar B_s$ mixing interplay with $B$ anomalies}

\author[a,b]{Luca Di Luzio\footnote{Talk given at the 10th International 
Workshop on the CKM 
Unitarity Triangle. Heidelberg University, September 17--21 2018.}}
\author[b,c,d]{Matthew Kirk}
\author[b]{Alexander Lenz}

\address[a]{Dipartimento di Fisica dell' Universit\`a di Pisa and INFN, Italy}
\address[b]{Institute for Particle Physics Phenomenology, Durham University, \\
DH1 3LE Durham, United Kingdom}
\address[c]{Dipartimento di Fisica, Universit\`a di Roma ``La Sapienza'', Piazzale Aldo Moro 2, 00185 Roma, Italy}
\address[d]{INFN Sezione di Roma, Piazzale Aldo Moro 2, 00185 Rome, Italy
\\ \ \\
luca.diluzio@pi.infn.it, matthew.kirk@roma1.infn.it, alexander.lenz@durham.ac.uk}

\begin{abstract}
After reviewing the theoretical uncertainties entering the Standard Model 
determination of the mass difference of the neutral $B_s$--$\bar B_s$ 
meson system, $\Delta M_s^{\rm SM}$, 
we discuss the implications of its updated value 
for new physics models addressing the 
experimental anomalies in semi-leptonic $B$ decays.
Using the most recent FLAG average of lattice results for the 
non-perturbative matrix elements and the CKM-fitter determination 
of $V_{cb}$ points to a $1.8 \, \sigma$ discrepancy in 
$\Delta M_s^{\rm SM} > \Delta M_s^{\rm exp}$. 
Extending the analysis in Ref.~\cite{DiLuzio:2017fdq} we show that 
the latter tension cannot be easily accommodated within single mediator models, 
whenever the same mediator is also responsible for the $b \to s \ell \ell$ 
anomalies. 
\end{abstract}


\end{frontmatter}

\clearpage

\section{Introduction}
\label{sec:intro}

While awaiting the LHCb Run-2 updates about the tantalizing 
hints of new physics (NP) in semi-leptonic $B$-meson 
decays \cite{Aaij:2014ora,Aaij:2017vbb,Aaij:2015esa,Khachatryan:2015isa,Lees:2015ymt,Wei:2009zv,Aaltonen:2011ja,Aaij:2015oid,Wehle:2016yoi,Sirunyan:2017dhj,Lees:2013uzd,Aaij:2015yra,Hirose:2016wfn} 
it would be natural to expect possible deviations from the Standard Model (SM) 
also in 4-quark and 4-lepton effective operators.
In fact, it is almost a theorem that a NP contribution say in 
$b \to s \ell \ell$ will eventually feed into a $(b s^\dag)^2$ operator. 
The latter are very well constrained by 
the measurement of the mass difference of the neutral $B_s$--$\bar B_s$ 
meson system, $\Delta M_s$, which provides a severe constraint  
for any NP model aiming at an explanation of the $B$-physics anomalies.  

For quite some time the SM value for $\Delta M_s$ 
was in perfect agreement with experimental results,
see e.g.~\cite{Artuso:2015swg,Lenz:2011ti}.
Taking however, the most recent lattice inputs, in particular the new average provided by the Flavour Lattice Averaging Group (FLAG) 
one gets a SM value considerably above the measurement. 
In this note, which is based on Ref.~\cite{DiLuzio:2017fdq}, 
we briefly review the SM prediction of $\Delta M_s$ 
and discuss its impact on NP models addressing the $B$ anomalies. 
We also complement Ref.~\cite{DiLuzio:2017fdq} with an analysis 
of simplified $Z'$ models featuring either complex couplings 
or general left-handed (LH) and right-handed (RH) chirality structures, having in mind the possibility 
of fitting simultaneously both the $b \to s \ell \ell$ anomalies and the 
$\Delta M_s$ tension. We conclude, however, that the two latter observables 
cannot be straightforwardly accommodated within a single-mediator simplified model.

\section{$\Delta M_s$ in the Standard Model}
\label{DeltaMsSM}

The mass difference of the mass eigenstates of the neutral $B_s$ mesons is given by
\begin{equation}
\Delta M_s \equiv M_H^s - M_L^s
=
2 \left|M_{12}^s\right| \, .
\label{DMdef}
\end{equation}
\begin{figure}[ht]
\centering
\includegraphics[width=0.9 \textwidth]{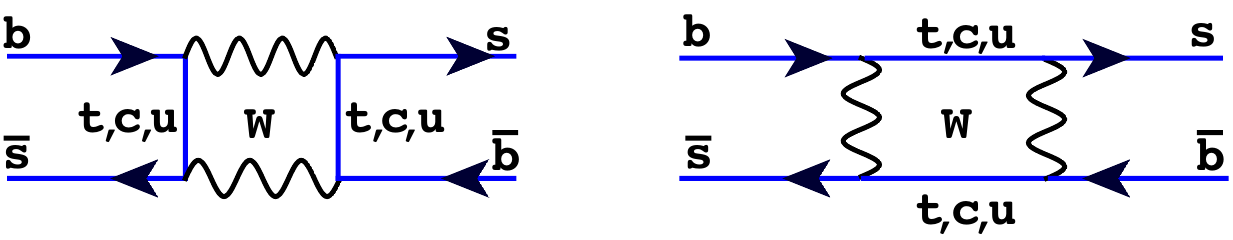}
\caption{\label{box} SM diagrams for the transition between $B_s$ and $\bar{B}_s$ mesons.
 The contribution of internal off-shell particles is denoted by $M_{12}^s$. 
}
\end{figure}

The calculation of the box diagrams in Fig.~\ref{box} gives the SM value for $M_{12}^s$,
see e.g.~\cite{Artuso:2015swg} for a brief review, and one gets
\begin{equation}
M_{12}^s = \frac{G_F^2}{12 \pi^2}
\lambda_t^2 M_W^2 S_0(x_t)
B f_{B_s}^2  M_{B_s} \hat{\eta }_B\, ,
\label{M12}
\end{equation}
with the Fermi constant $G_F$, the masses of the $W$ boson,
$M_W$, and of the $B_s$ meson, $M_{B_s}$. 
Using CKM unitarity one finds only one contributing CKM structure 
$ \lambda_t = V_{ts}^* V^{}_{tb}$.
The CKM elements are the only place in Eq.~(\ref{M12}) where an imaginary part can arise.
The result of the 1-loop diagrams given in Fig.~\ref{box}
is denoted by the Inami-Lim function \cite{Inami:1980fz}
$ S_0(x_t= (\bar{m}_t(\bar{m}_t))^2/M_W^2) \approx 2.36853$, where $\bar{m}_t(\bar{m}_t)$ is
the $\overline{\rm MS}$-mass \cite{Bardeen:1978yd} of the top quark.
Perturbative 2-loop QCD corrections are encoded in the factor
$\hat{\eta }_B \approx 0.83798$ \cite{Buras:1990fn}.
In the SM calculation of $M_{12}^s$ one four quark $\Delta B=2$ operator
arises
\begin{equation}
Q = \bar s^\alpha \gamma_\mu (1- \gamma_5) b^\alpha
        \times
        \bar s^\beta  \gamma^\mu (1- \gamma_5) b^\beta \, .
      \label{Q}
\end{equation}
The hadronic matrix element of this operator is parametrised in terms of a decay constant $f_{B_s}$ and a bag parameter
$B$:
\begin{equation}
\langle Q \rangle \equiv \langle B_s^0|Q |\bar{B}_s^0 \rangle
       =
      \frac{8}{3}  M_{B_s}^2  f_{B_s}^2 B (\mu) \, ,
\label{ME1}
\end{equation}
We also indicated the renormalisation scale dependence
of the bag parameter; in our analysis we take $\mu = \bar{m}_b(\bar{m}_b)$. 
\\
Sometimes a different notation for the QCD corrections and the bag parameter
is used in the literature
(e.g.~by FLAG \cite{Aoki:2016frl}),
$(\eta_B, \hat{B})$ instead of $(\hat{\eta}_B, B)$ 
with $\hat{\eta}_B B \equiv \eta_B \hat{B}$ and $\hat{B} = 1.51926 \, B$. 
The parameter $\hat{B}$ has the advantage
of being renormalisation scale and scheme independent.
\\
A commonly used SM prediction of $\Delta M_s$ was given by \cite{Artuso:2015swg}
\begin{align}
\label{DeltaMold2015}
\Delta M_s^{\rm SM,\, 2015} &= \left(18.3 \pm 2.7 \right)
\, \mbox{ps}^{-1} \; , 
\end{align}
that agreed very well with the experimental measurement \cite{Amhis:2016xyh}
\begin{equation}
\Delta M_s^{\rm Exp} = \left(17.757 \pm 0.021 \right)\; \mbox{ps}^{-1} \; .
\label{DeltaMExp}
\end{equation}
In 2016 Fermilab/MILC presented a new calculation \cite{Bazavov:2016nty}, 
which gave considerably larger values for
the non-perturbative parameter, resulting in values around 20 ps$^{-1}$ 
for the mass difference \cite{Bazavov:2016nty,Blanke:2016bhf,Jubb:2016mvq,Buras:2016dxz,Altmannshofer:2017uvs} 
and being thus larger than the experimental measurement. 
An independent confirmation of these large values would of course be desirable; 
a first step in that direction has been done by the HQET sum rule
calculation of \cite{Kirk:2017juj}.
In that work they calculate only the bag parameters for $B_d$-mixing, however these should be close to those for $B_s$-mixing -- a preliminary result for the $B_s$-mixing parameters was presented at CKM2018 \cite{RauhCKM2018}.
Their results for the bag parameters agree (within uncertainties) with Fermilab/MILC.

Using the most recent numerical inputs we predict the mass difference of the neutral $B_s$ mesons to be \cite{DiLuzio:2017fdq}\footnote{A more conservative determination of the SM value of 
the mass difference using only tree-level inputs for the CKM parameters 
is $\Delta M_s^\text{SM, 2017 (tree)} = (19.9 \pm 1.5) \; \text{ps}^{-1}$ \cite{DiLuzio:2017fdq}.}
\begin{equation}
\Delta M_s^{\rm SM,\, 2017} = \left(20.01 \pm 1.25 \right) \; \mbox{ps}^{-1} \; .
\label{DeltaM2017}
\end{equation}
Here the dominant uncertainty still comes  from the lattice predictions for the non-perturbative
parameters  $B$ and $ f_{B_s}$, giving a relative error of $6 \%$.
The uncertainty in the CKM elements (determined assuming unitarity of the CKM matrix) 
contributes $2\%$ to the error budget. 
Other uncertainties can be safely neglected at the current stage.
The new central value for the mass difference in Eq.~(\ref{DeltaM2017}) is 1.8 $\sigma$ above 
the experimental one given in Eq.~(\ref{DeltaMExp}). 
This difference has profound implications for NP models 
that predict sizeable positive contributions to $\Delta M_s$. 
The new value for the SM prediction depends strongly on the
non-perturbative input as well as the values of the CKM elements (in particular the element $V_{cb}$). 
We use the averages that are provided by the lattice community 
(web-update of FLAG \cite{Aoki:2016frl}) 
and by the CKMfitter group (web-update of \cite{Charles:2004jd} -- 
similar values can be taken from the UTfit group \cite{Bona:2006ah}). 
For further details we refer the reader to Ref.~\cite{DiLuzio:2017fdq}.

\section{$\Delta M_s$ beyond the Standard Model}
\label{BmixingBSM}
To determine the allowed space for NP effects in $B_s$-mixing we compare the experimental
measurement of the mass difference with the prediction in the SM plus NP: 
\begin{equation}
\Delta M_s^{\rm Exp} = 2 \left| M_{12}^{\rm SM} + M_{12}^{\rm NP} \right|
= \Delta M_s^{\rm SM} \left| 1 +  \frac{M_{12}^{\rm NP}}{M_{12}^{\rm SM}}\right|  \, .
\label{MsExpvsNP}
\end{equation}
In the following, we will assume that NP effects do not involve sizeable shifts in the CKM elements.

A simple estimate shows that the improvement of the SM prediction from 
\eq{DeltaMold2015} to 
\eq{DeltaM2017} can have a drastic impact on the size of the allowed NP effects on $B_s$-mixing. 
For a generic NP model we can parametrise 
\begin{equation}
\label{LambdaNP}
\frac{\Delta M_s^{\rm Exp}}{\Delta M_s^{\rm SM}} = \abs{1 + \frac{\kappa}{\Lambda^2_{\rm NP}}} \, ,
\end{equation}
where $\Lambda_{\rm NP}$ denotes the mass scale of the NP mediator and $\kappa$ is a dimensionful quantity 
which encodes NP couplings and the SM contribution. If $\kappa > 0$ 
(this is often the case in many NP scenarios for $B$ anomalies), 
and since 
$\Delta M_s^{\rm SM} > \Delta M_s^{\rm Exp}$, 
the 2$\sigma$ bound on $\Lambda_{\rm NP}$ scales like
\begin{equation}
\label{scalingNP}
\frac{\Lambda^{\rm 2017}_{\rm NP}}{\Lambda^{\rm 2015}_{\rm NP}} 
= \sqrt{
\frac{\frac{\Delta M_s^{\rm Exp}}{\left(\Delta M_s^{\rm SM} - 2 \delta\Delta M_s^{\rm SM}\right)^{2015}} - 1}
{\frac{\Delta M_s^{\rm Exp}}{\left(\Delta M_s^{\rm SM} - 2 \delta\Delta M_s^{\rm SM}\right)^{2017}} - 1}}
\simeq 5.2 \, ,
\end{equation}
where $\delta \Delta M_s^{\rm SM}$ denotes the 1$\sigma$ error of the SM prediction. 
Hence, in models where $\kappa > 0$, the limit on the mass of the NP mediators is strengthened by a factor 5. 
On the other hand, if the tension between the SM prediction and $\Delta M_s^{\rm Exp}$ 
increases in the future, 
a NP contribution with $\kappa < 0$ would be required in order to accommodate the 
discrepancy. 

A typical example where $\kappa > 0$ is that of a purely LH vector-current operator, 
which arises from the exchange 
of a single mediator featuring real couplings, 
cf.~\sect{BmixvsBanom}.
In such a case, the short-distance contribution to $B_s$-mixing is described by the effective Lagrangian 
\begin{equation}
\label{LNPDB2}
\mathcal{L}^{\rm NP}_{\Delta B = 2} = - \frac{4 G_F}{\sqrt{2}} \left( V^{}_{tb} V^*_{ts} \right)^2 
\left[ C^{LL}_{bs} \left( \bar s_L \gamma_\mu b_L \right)^2 + \text{h.c.} \right] \, ,
\end{equation}
where $C^{LL}_{bs}$ is a Wilson coefficient to be matched with some ultraviolet (UV) model. 
This coefficient enters Eq.~(\ref{MsExpvsNP}) as  
\begin{equation}
\frac{\Delta M_s^{\rm Exp}}{\Delta M_s^{\rm SM}}
= \abs{1 + \frac{C^{LL}_{bs}}{R^{\rm loop}_{\rm SM}}}
\, , 
\end{equation}
where 
\begin{equation}
R^{\rm loop}_{\rm SM} 
= \frac{\sqrt{2} G_F M_W^2 \hat\eta_B S_0(x_t)}{16 \pi^2} = 1.3397 \times 10^{-3} \, .
\end{equation}

In the following,
we will show how the updated bound from $\Delta M_s$ impacts the parameter space 
of simplified models (with $\kappa > 0$) put forth for the explanation of the recent discrepancies in 
semi-leptonic B-physics data (\sect{BmixvsBanom}) 
and then discuss the feasability of some $\kappa < 0$ scenarios (\sect{NPmodelskl0}).

\subsection{Impact of $B_s$-mixing on NP models for B anomalies}
\label{BmixvsBanom}

A useful application of the refined SM prediction in Eq.~(\ref{DeltaM2017}) 
is in the context of the recent hints of LFU violation in semi-leptonic $B$-meson decays. 
Focussing on neutral current anomalies, the main observables are 
the LFU violating ratios $R_{K^{(*)}} \equiv \mathcal{B}(B \to K^{(*)} \mu^+ \mu^-) / \mathcal{B}(B \to K^{(*)} e^+ e^-)$ 
\cite{Aaij:2014ora,Aaij:2017vbb}, together with the angular distributions of $B \to K^{(*)} \mu^+ \mu^-$ \cite{Aaij:2015esa,Khachatryan:2015isa,Lees:2015ymt,Wei:2009zv,Aaltonen:2011ja,Aaij:2015oid,Abdesselam:2016llu,Wehle:2016yoi,ATLAS:2017dlm,Sirunyan:2017dhj}
and the branching ratios of hadronic $b \to s \mu^+ \mu^-$ decays \cite{Aaij:2014pli,Aaij:2015esa,Khachatryan:2015isa}. 
As hinted by various recent global fits 
\cite{Capdevila:2017bsm,Altmannshofer:2017yso,DAmico:2017mtc,Alok:2017sui,Geng:2017svp,Ciuchini:2017mik,Hiller:2017bzc}, 
and in order to simplify a bit the discussion, 
we assume NP contributions only in purely LH vector currents involving muons. 
The effective Lagrangian for semi-leptonic $b \to s \mu^+ \mu^-$ transitions contains the terms
\begin{equation}
\label{Leffbsmumu}
\mathcal{L}^{\rm NP}_{b \to s \mu \mu} \supset \frac{4 G_F}{\sqrt{2}} V^{}_{tb} V^*_{ts} \left( \delta C^\mu_9 O^\mu_9 + \delta C^\mu_{10} O^\mu_{10} \right) + \text{h.c.} \, ,
\end{equation}
with 
\begin{align}
O^\mu_9 &= \frac{\alpha}{4 \pi} (\bar s_L \gamma_\mu b_L) (\bar \mu \gamma^\mu \mu) \, , \\
O^\mu_{10} &= \frac{\alpha}{4 \pi} (\bar s_L \gamma_\mu b_L) (\bar \mu \gamma^\mu \gamma_5 \mu) \, . 
\end{align}
Assuming purely LH currents
and real Wilson coefficients the 
best-fit of $R_{K}$ and $R_{K^*}$ yields 
(from e.g.~\cite{Altmannshofer:2017yso}):
$\Re (\delta C^\mu_9) = - \Re (\delta C^\mu_{10}) \in [-0.81, -0.48]$ ($[-1.00, -0.32]$) at 1$\sigma $ (2$\sigma$). 
Adding also the data on $B \to K^{(*)} \mu^+\mu^-$ angular distributions and 
other $b \to s \mu^+ \mu^-$ observables improves the statistical significance of the fit, 
but does not necessarily implies larger deviations of $\Re (\delta C^\mu_9)$ from zero (see e.g.~\cite{Capdevila:2017bsm}). 
For the results first presented in Ref.~\cite{DiLuzio:2017fdq}, we will stick only to the $R_{K}$ and $R_{K^*}$ observables and denote this benchmark as ``$R_{K^{(*)}}$'', 
while for new results we present here a wider range of observables is used (denoted by ``$b \to s \ell \ell $'').

\subsubsection{Z'}
A paradigmatical NP model for explaining the $B$ anomalies in neutral currents 
is that of a $Z'$ dominantly coupled via LH currents.  
Here, we focus only on the part of the Lagrangian relevant for $b \to s \mu^+ \mu^-$ transitions and $B_s$-mixing, namely 
\begin{equation}
\label{LZp}
\mathcal{L}_{Z'} = 
\frac{1}{2} M^2_{Z'} (Z'_\mu)^2 +
\left( \lambda^Q_{ij} \, \bar d_L^i \gamma^\mu d_L^j + \lambda^L_{\alpha\beta} \, \bar \ell_L^\alpha \gamma^\mu \ell_L^\beta \right) Z'_\mu 
\, , 
\end{equation}
where $d^i$ and $\ell^\alpha$ denote down-quark and charged-lepton mass eigenstates, and 
$\lambda^{Q,L}$ are 
hermitian matrices in flavour space. 
Of course, any full-fledged (i.e.~$SU(2)_L \times U(1)_Y$ 
gauge invariant and anomaly free) $Z'$ model attempting an explanation of $R_{K^{(*)}}$ 
via LH currents can be mapped into Eq.~(\ref{LZp}). 
After integrating out the $Z'$ at tree level, we obtain the 
effective Lagrangian
\begin{align}
\mathcal{L}_{Z'}^{\rm eff} &= -\frac{1}{2 M^2_{Z'}} \left( \lambda^Q_{ij} \, \bar d_L^i \gamma_\mu d_L^j 
+ \lambda^L_{\alpha\beta} \, \bar \ell_L^\alpha \gamma_\mu \ell_L^\beta \right)^2  \\
&\supset -\frac{1}{2 M^2_{Z'}} 
\left[ 
(\lambda^Q_{23})^2 \left( \bar s_L \gamma_\mu b_L \right)^2 + 2 \lambda^Q_{23} \lambda^L_{22} (\bar s_L \gamma_\mu b_L) (\bar \mu_L \gamma^\mu \mu_L)
+ \text{h.c.}
\right]
\, . \nonumber
\end{align}
Matching with Eq.~(\ref{Leffbsmumu}) and (\ref{LNPDB2}) we get 
\begin{equation}
\delta C^\mu_9 = -\delta C^\mu_{10} = - \frac{\pi}{\sqrt{2} G_F M^2_{Z'} \alpha} \left( \frac{\lambda^Q_{23} \lambda^L_{22}}{V^{}_{tb} V^*_{ts}} \right) \, ,
\end{equation}
and 
\begin{equation} 
\label{CbsZp}
C^{LL}_{bs} = \frac{\eta^{LL}(M_{Z'})}{4 \sqrt{2} G_F M^2_{Z'}} \left( \frac{\lambda^Q_{23}}{V^{}_{tb} V^*_{ts}} \right)^2
\, , 
\end{equation}
where $\eta^{LL}(M_{Z'})$ 
encodes the running down to the bottom mass scale using NLO anomalous dimensions \cite{Ciuchini:1997bw,Buras:2000if}. 
E.g.~for $M_{Z'} \in [1,10]$ TeV we find $\eta^{LL}(M_{Z'}) \in [0.79,0.75]$.   

Here, we first consider the case of a real coupling $\lambda^Q_{23}$, so that 
$C^{LL}_{bs} > 0$ and $\delta C^\mu_9 = -\delta C^\mu_{10}$ 
is also real. This assumption follows the standard approach of nearly all the 
groups performing global fits 
\cite{Altmannshofer:2017fio,Ciuchini:2017mik,Geng:2017svp,Capdevila:2017bsm,Altmannshofer:2017yso,DAmico:2017mtc,Alok:2017sui}.  
The case of complex $Z'$ couplings will be considered in \sect{complex_couplings}.

\begin{figure}[htb!]
\center
\includegraphics[width=.6\textwidth]{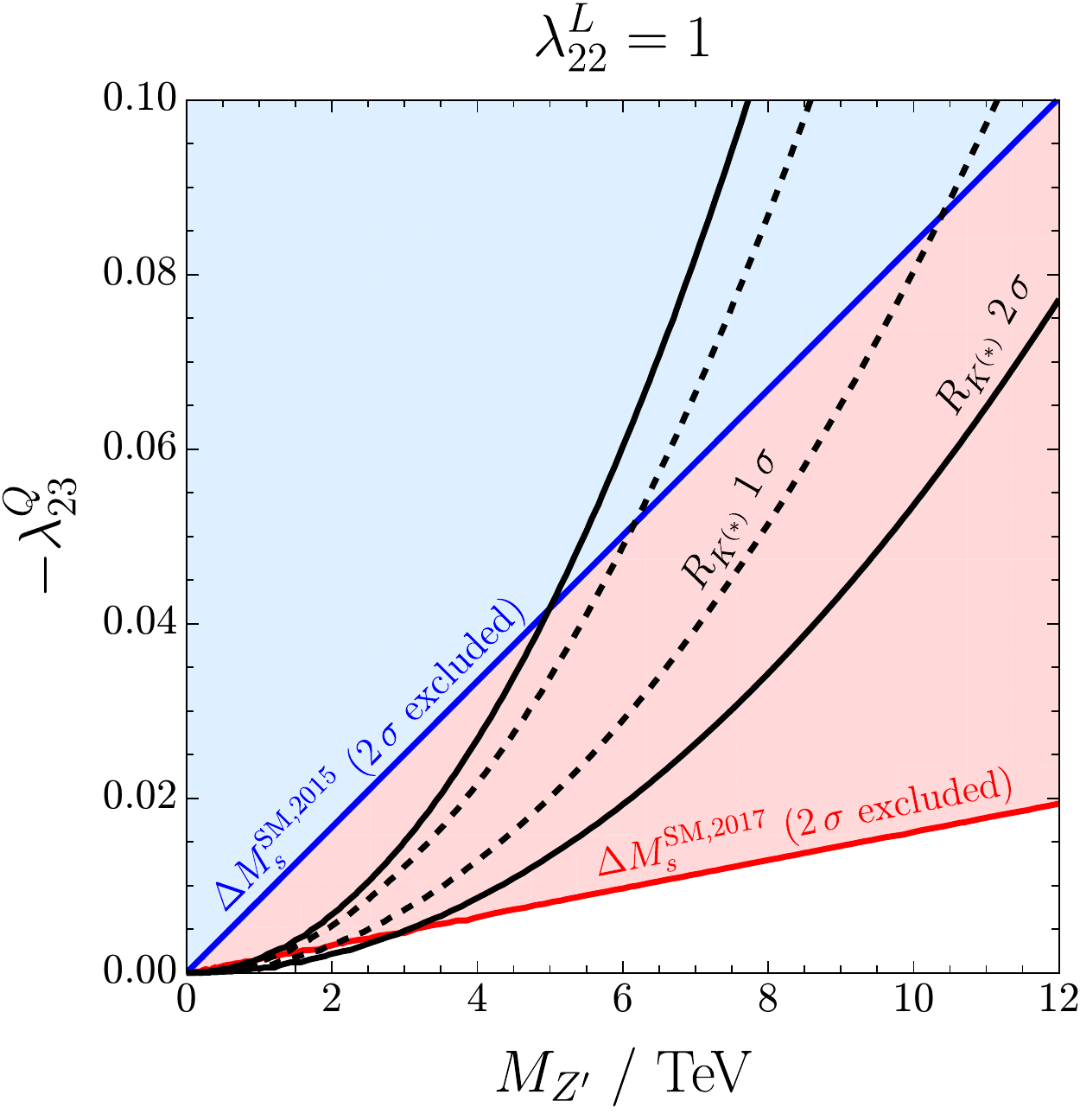} 
\caption{\label{fig:BsmixvsRK}
Bounds from $B_s$-mixing on the parameter space of the 
simplified $Z'$ model of \eq{LZp}, for real $\lambda^Q_{23}$ and $\lambda^L_{22} =1$. 
The blue and red shaded areas correspond respectively to the 2$\sigma$ exclusions from $\Delta M_s^{\rm SM,\,2015}$ and 
$\Delta M_s^{\rm SM,\,2017}$, while the solid (dashed) black curves encompass the 2$\sigma$ (1$\sigma$) 
best-fit region from $R_{K^{(*)}}$. 
}
\end{figure}

The impact of the improved SM calculation of $B_s$-mixing on the parameter space of the $Z'$ explanation of 
$R_{K^{(*)}}$ is displayed in Fig.~\ref{fig:BsmixvsRK}, for the reference value $\lambda^L_{22} =1$.\footnote{For 
$m_{Z'} \lesssim 1$ TeV the coupling $\lambda^L_{22}$ is bounded by the $Z\to 4\mu$ measurement at LHC and 
by neutrino trident production \cite{Altmannshofer:2014pba}. 
See for instance Fig.~1 in \cite{Falkowski:2018dsl} for a recent analysis.}
Note that the old SM determination, $\Delta M_s^{\rm SM,\,2015}$, allowed for $M_Z'$ as heavy as 
$\approx 10$ TeV in order to explain $R_{K^{(*)}}$ at 1$\sigma$. 
In contrast, $\Delta M_s^{\rm SM,\,2017}$ implies now $M_Z' \lesssim 2$ TeV.
Even for $\lambda^L_{22} =\sqrt{4 \pi}$, which saturates the perturbative unitarity bound \cite{DiLuzio:2017chi,DiLuzio:2016sur}, 
we find that the updated limit from $B_s$-mixing requires 
$M_Z' \lesssim 8$ TeV for the 1$\sigma$ explanation of $R_{K^{(*)}}$. 
Whether a few TeV $Z'$ is ruled out or not by direct searches at LHC depends however on the details of the $Z'$ model.  
For instance, the stringent constraints from di-lepton searches \cite{ATLAS:2017wce} are tamed in models where the $Z'$ 
couples mainly to third generation fermions (as e.g.~in \cite{Alonso:2017uky}).  
This notwithstanding, the updated limit from $B_s$-mixing cuts dramatically into the parameter space 
of the $Z'$ explanation of the $b \to s \mu^+ \mu^-$ anomalies.

\section{Model building directions for $\Delta M_s^{\rm NP}<0$}
\label{NPmodelskl0}

Given that $\Delta M_s^{\rm SM} > \Delta M_s^{\rm exp}$ at about 2$\sigma$, 
it is worth to investigate possible ways to obtain a negative NP contribution 
to $\Delta M_s$, thus relaxing the tension between the SM and the experimental measurement. 

Sticking to the simplified model of \sect{BmixvsBanom} ($Z'$ coupled only to LH currents), an obvious solution in order to achieve $C^{LL}_{bs}<0$ 
is to allow for complex couplings (cf.~\eq{CbsZp}).
For instance, in $Z'$ models this could happen as a consequence of fermion 
mixing if the $Z'$ does not couple universally in the gauge-current basis 
(see e.g.~\cite{Gauld:2013qja}). 
Extra phases in the couplings are constrained by CP-violating observables, which we will discuss in \sect{complex_couplings}.

An alternative way to achieve a negative contribution for $\Delta M_s^{\rm NP}$ is to go beyond 
the simplified models of \sect{BmixvsBanom} and contemplate generalised chirality structures. 
Let us consider for definiteness the case of a $Z'$ coupled both to LH and RH down-quark currents
\begin{equation}
\mathcal{L}_{Z'} \supset \frac{1}{2} M^2_{Z'} (Z'_\mu)^2 + \left( \lambda^Q_{ij} \, \bar d_L^i \gamma^\mu d_L^j + 
\lambda^d_{ij} \, \bar d_R^i \gamma^\mu d_R^j \right) Z'_\mu \, .
\end{equation}
Upon integrating out the $Z'$ one obtains 
\begin{align}
\mathcal{L}_{Z'}^{\rm eff} 
& \supset -\frac{1}{2 M^2_{Z'}} 
 \left[ 
(\lambda^Q_{23})^2 \left( \bar s_L \gamma_\mu b_L \right)^2 
+ (\lambda^d_{23})^2 \left( \bar s_R \gamma_\mu b_R \right)^2 \right. \nonumber \\
& \left. \qquad \qquad \qquad + 2 \lambda^Q_{23} \lambda^d_{23} (\bar s_L \gamma_\mu b_L) (\bar s_R \gamma_\mu b_R)
+ \text{h.c.}
\right]
\, . 
\end{align}
The LR vector operator can clearly have any sign, even for real couplings, and we take up this possibility in \sect{rh_quark_coupling}.

\subsection{Complex Couplings}
\label{complex_couplings}

In this section we consider the case of complex couplings, first from a model independent 
perspective (\sect{complex_couplings_modelind}) and then in a specific $Z'$ model (\sect{complex_couplings_Zp}).  

\subsubsection{Fit to complex $\delta C^\mu_9$}
\label{complex_couplings_modelind}

So far the focus of global fits has been on NP coefficients with the same phase as the SM contributions (which are essentially real as only a very small phase is generated by $\operatorname{Arg} (V^{}_{tb} V^*_{ts}) = -3.12 \approx -179^\circ$), 
barring however few exceptions \cite{Alok:2017jgr,Alda:2018mfy}.
Here, we extend the study in Ref.~\cite{DiLuzio:2017fdq} by performing our own fit 
to a specific scenario where NP only arises in $\delta C_9^\mu = -\delta C_{10}^\mu$, 
using \texttt{flavio} \cite{Straub:2018kue}.
\begin{figure}[htb!]
\center
\includegraphics[width=.6\textwidth]{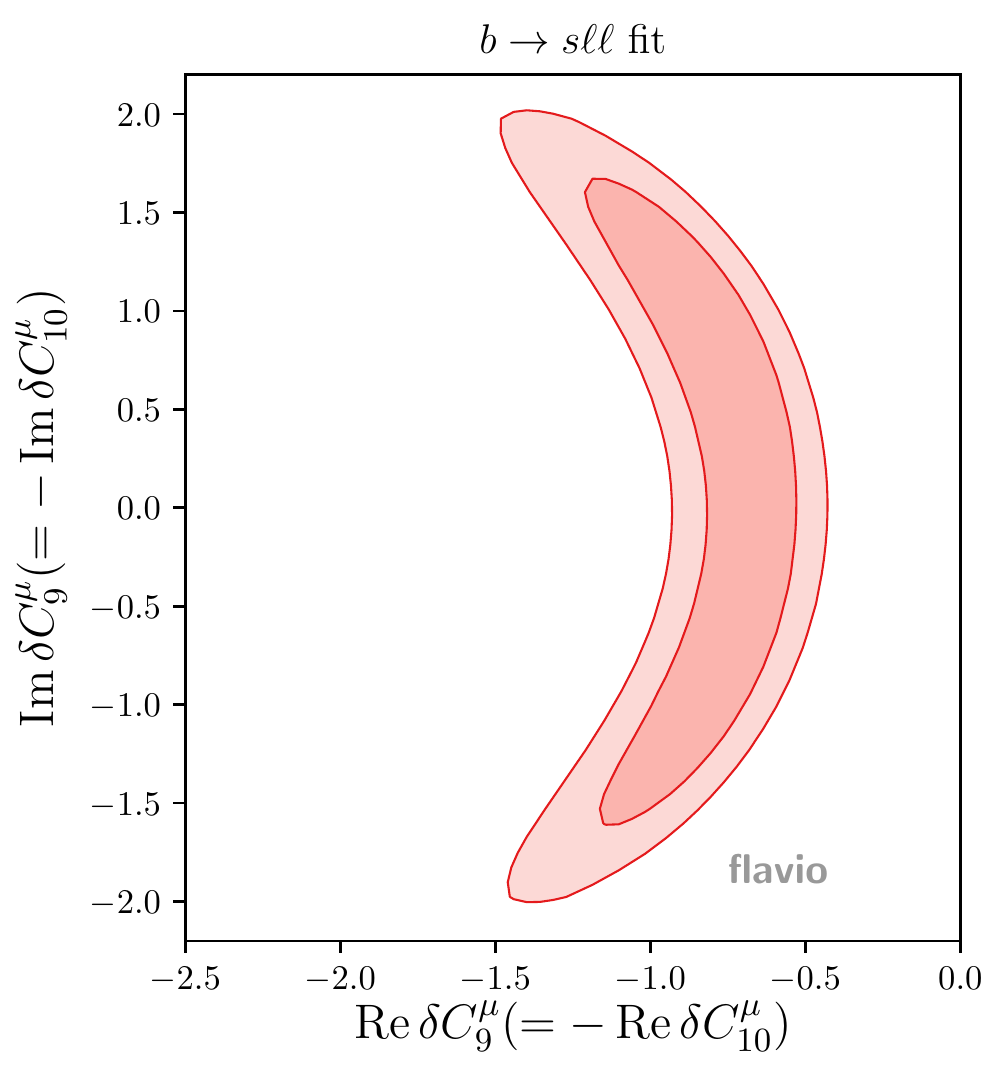} 
\caption{Fit to complex $\delta C_9^\mu$ couplings, with $\delta C_9^\mu = - \delta C_{10}^\mu$. 
The darker (lighter) shaded area shows the 1$\sigma$ (2$\sigma$) region favoured by the fit.
The employed set of observables is taken from \cite{Altmannshofer:2017fio} and is denoted 
collectively by ``$b \to s \ell \ell$''.}
\label{fig:complex_c9_c10}
\end{figure}
The result is shown in \fig{fig:complex_c9_c10} -- 
we see that while there is a relatively narrow range for the real part to explain the flavour anomalies, 
the imaginary part has much more freedom.
(The shape of the allowed region in the complex $\delta C_9^\mu$ space 
matches that found by \cite{Alok:2017jgr}.)
This can be qualitatively understood from the fact that 
the imaginary part only arise quadratically in the expressions for $R_{K^{(*)}}$ 
since the leading interference term with the SM amplitude is real. Hence the imaginary part 
is relatively unconstrained by the fit unless $\Im \delta C_9^\mu \gtrsim \Re \delta C_9$.

\subsubsection{Complex $Z^\prime$}
\label{complex_couplings_Zp}

Once we allow the $Z'$ quark coupling to be complex, there are extra constraints to be considered, in the form of CP-violating observables that arise from $B_s$-mixing.
The most relevant here is the mixing-induced CP asymmetry \cite{Lenz:2006hd,Artuso:2015swg}, arising from interference between B meson mixing and decay. 
The semi-leptonic CP asymmetries for flavour-specific decays, $a^s_{\rm sl}$, 
are not competitive here since the experimental errors are still too large \cite{Artuso:2015swg}.
Defining 
\begin{equation} 
\phi_{\Delta} = \text{Arg} \left( 1 + \frac{C^{LL}_{bs}}{R^{\rm loop}_{\rm SM}} \right) \, ,
\end{equation}
the mixing-induced CP asymmetry is given by
\begin{equation}
A^{\rm mix}_{\rm CP} (B_s \to J/\psi \phi) = \sin{\left( \phi_\Delta - 2 \beta_s \right)} \, ,
\end{equation}
where $A^{\rm mix}_{\rm CP} = -0.021 \pm 0.031$ \cite{Amhis:2016xyh}, $\beta_s = 0.01852 \pm 0.00032$ \cite{Charles:2004jd}, 
and we neglected penguin contributions \cite{Artuso:2015swg}.

\begin{figure}[tb]
\center
\includegraphics[width=.6\textwidth]{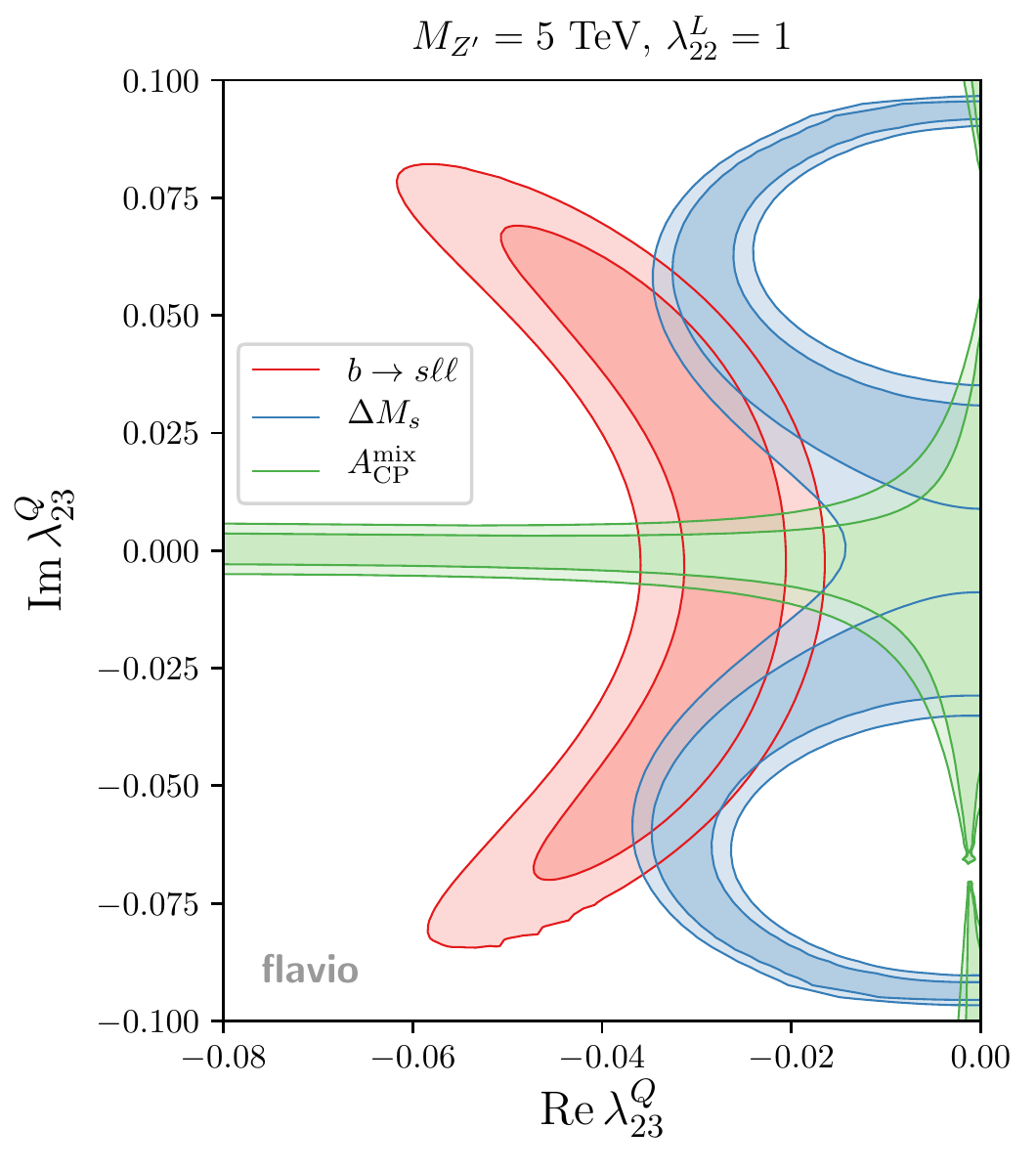} 
\caption{Fit to complex $Z'$ couplings. The darker (lighter) shaded regions show the 1$\sigma$ (2$\sigma$) allowed regions respectively.}
\label{fig:complex_Zprime}
\end{figure}

Including this extra observable in our fit, 
we display our results in \fig{fig:complex_Zprime}, 
for the reference values $M_Z' = 5$ TeV and $\lambda^L_{22} = 1$.
While there are regions in which both $b \to s \ell \ell$ and $\Delta M_s$ can be 
accommodated at $1 \sigma$, 
the additional constraint from $A^{\rm mix}_{\rm CP}$ 
precludes this possibility by setting a strong a limit on the imaginary part of the $Z'$ coupling.

\subsection{Fit with RH quark coupling}
\label{rh_quark_coupling}

As discussed above, if we extend the minimal model to include both LH and RH down-quark currents, 
there arises an interference term in $\Delta M_s$ with arbitrary sign.
Moreover, since this term gets enhanced by renormalisation-group effects compared to LL and RR vector operators \cite{Buras:2001ra}, it can easily dominate the contribution to $\Delta M_s^{\rm NP}$. 
However, while there are no extra constraints to be taken into account as for the case of a complex coupling, 
this scenario brings in its own problem -- namely that the contribution to $R_{K^{(*)}}$ via RH quark currents 
must be sizable. 
Current global fits disfavour a purely RH quark current, as this breaks the experimentally observed relation $R_{K} \approx R_{K^{*}}$ (see e.g.~\cite{DAmico:2017mtc} for further details).
The question then is whether a combined explanation of $R_{K^{(*)}}$ and $\Delta M_s$ is possible within the framework of current experimental results.

\begin{figure}[tbh]
\center
\includegraphics[width=.6\textwidth]{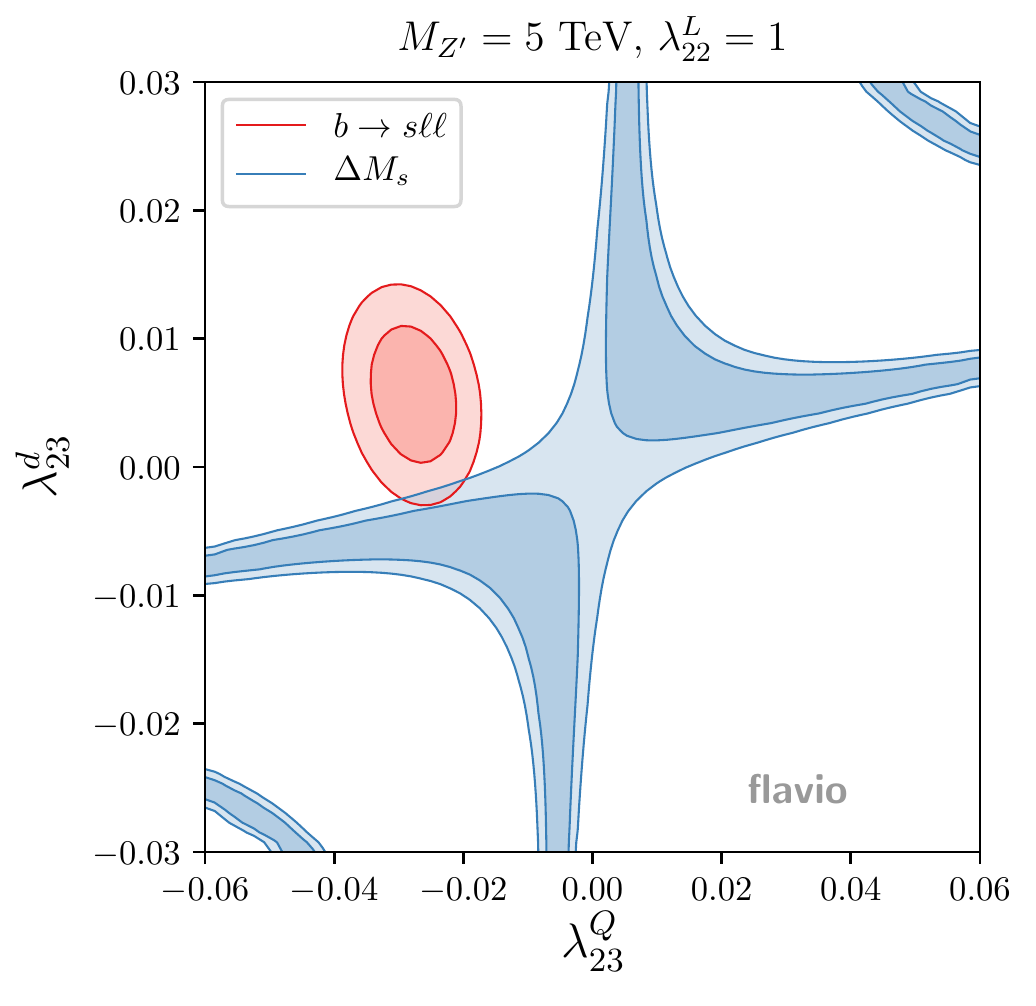} 
\caption{Fit to $Z'$ with LH and RH quark couplings. The darker (lighter) shaded regions show the 1$\sigma$ (2$\sigma$) allowed regions respectively.}
\label{fig:LR_Zprime}
\end{figure}

Our results are shown in \fig{fig:LR_Zprime} -- 
while a negative contribution to $\Delta M_s$ favours the LH and RH quark couplings to have the 
same sign,\footnote{Note that the matrix element of the vector LR operator is negative, 
while that of the LL and RR operators is positive.}  
the small region favoured by the semi-leptonic $B$ anomalies 
has no overlap with the $\Delta M_s$ region at 1$\sigma$.

\section{Conclusions}
\label{conclusions}

In this note, we have restated our update \cite{DiLuzio:2017fdq} of the SM prediction for the 
$B_s$-mixing observable \(\Delta M_s\) (\eq{DeltaM2017}) using the most recent values for the input parameters, 
in particular the latest lattice results from FLAG.
Our update shifts the central value of the SM theory prediction upwards and implies 
a 1.8$\sigma$ discrepancy from the SM.

We further discussed an important application of the \(\Delta M_s\) update for 
NP models aimed at explaining the recent anomalies in semi-leptonic $B$ decays.
The latter typically predict a positive shift in the NP contribution to 
\(\Delta M_s\), thus making the discrepancy with respect to the experimental value even worse.   
As a generic result we have shown that, whenever the NP contribution to \(\Delta M_s\) is positive, 
the limit on the mass of the NP mediator that must be invoked in order to explain the anomalies 
is strengthened by a factor of five (for a fixed coupling) compared to using the 2015 SM calculation 
for $\Delta M_s$ -- a representative example of a simplified model of this type is a \(Z^\prime\) 
featuring purely LH and real couplings in order to accommodate \(R_{K^{(*)}}\). 
The improvement in the upper bound on the $Z'$ mass is shown in \fig{fig:BsmixvsRK}.

Here we extended our study \cite{DiLuzio:2017fdq} 
to investigate potential ``loopholes'' to those results, 
whereby a \emph{negative} contribution to \(\Delta M_s\) could arise that would lessen the tension in $B_s$-mixing 
while still providing a good fit to the currently observed $B$ anomalies.
Two cases were investigated -- one where we allowed the quark coupling in our minimal 
$Z'$ model to be complex and another where we extended the minimal model with $Z'$ couplings to RH down quarks. 

For the case of complex coupling, we showed that despite the fact that a relatively large imaginary part for 
$\delta C_9^\mu$ is compatible with the $b \to s \ell \ell$ data, 
any extra phase present in $B_s$-mixing is tightly constrained by the measurement of $A^\text{mix}_\text{CP}$ 
and this prevents an improvement of the overall fit (see \fig{fig:complex_Zprime}).

In the other extended case study, the results are again negative.
While it is known that adding a RH quark coupling is disfavoured by the $R_{K^{(*)}}$ fit 
(assuming NP in muons), 
it is also true that chirality-mixed LR vector operators give an RG enhanced contribution to $B_s$-mixing.
However, as we see from \fig{fig:LR_Zprime} the fit to $\Delta M_s$ and $b \to s \ell \ell$ data 
favours respectively the same and opposite sign combination for the $Z'$ couplings 
to LH and RH quarks.

Although other ways to accommodate $\Delta M_s$ together with the $B$ anomalies 
could certainly exist,\footnote{Here we mention two notable possibilities: 
$i)$ sticking only to $R_{K}$ and $R_{K^*}$, these can be accommodated via NP 
in electrons featuring sizeable contributions from RH quark currents, 
thus allowing also for negative contributions to $\Delta M_s$ (see e.g.~\cite{Carmona:2017fsn}) and 
$ii)$ as pointed out in \cite{Bordone:2018nbg}, in UV complete models 
of the vector leptoquark $U_\mu \sim (3,1,2/3)$ \cite{DiLuzio:2017vat,Calibbi:2017qbu,Bordone:2017bld,Barbieri:2017tuq,Blanke:2018sro,DiLuzio:2018zxy} addressing both $R_{D^{(*)}}$ and $R_{K^{(*)}}$, 
the fermion couplings of extra $Z'$/$G'$ states not directly related to the anomalies can naturally have 
a large phase in order to accommodate a negative $\Delta M_s$, without 
being in tension with CP violating observables.} 
we conclude that the simplest possibility of a single-mediator simplified model is disfavoured.  

We finally reiterate the importance of an independent confirmation 
of the FNAL/MILC lattice result for the four-quark matrix elements, 
given the central role of $B_s$-mixing in constraining 
NP models for $B$ anomalies.

\section*{Acknowledgments}
LDL would like to thank the organizers of CKM2018 for the kind invitation 
and for setting up a great atmosphere in Heidelberg.  
This work was supported by the STFC through the IPPP grant. 
The work of LDL is supported by the ERC grant NEO-NAT.


\bibliographystyle{elsarticle-num.bst}
\bibliography{bibliography}
\biboptions{sort&compress}

\end{document}